# Saliency Attention and Semantic Similarity-Driven Adversarial Perturbation

Hetvi Waghela[1][0009-004-2562-7632], Jaydip Sen[2][0000-0002-4120-8700] and Sneha Rakshit[3]

Praxis Business School, Kolkata 700104, INDIA
[1]waghelahetvi7@gmail.com, [2]jaydip.sen@acm.org,
[3]srakshit149@gmail.com

**Abstract.** In this paper, we introduce an enhanced textual adversarial attack method, known as Saliency Attention and Semantic Similarity driven adversarial Perturbation (SASSP). The proposed scheme is designed to improve the effectiveness of contextual perturbations by integrating saliency, attention, and semantic similarity. Traditional adversarial attack methods often struggle to maintain semantic consistency and coherence while effectively deceiving target models. Our proposed approach addresses these challenges by incorporating a three-pronged strategy for word selection and perturbation. First, we utilize a saliency-based word selection to prioritize words for modification based on their importance to the model's prediction. Second, attention mechanisms are employed to focus perturbations on contextually significant words, enhancing the attack's efficacy. Finally, an advanced semantic similarity-checking method is employed that includes embedding-based similarity and paraphrase detection. By leveraging models like Sentence-BERT for embedding similarity and fine-tuned paraphrase detection models from the Sentence Transformers library, the scheme ensures that the perturbed text remains contextually appropriate and semantically consistent with the original. Empirical evaluations demonstrate that SASSP generates adversarial examples that not only maintain high semantic fidelity but also effectively deceive state-of-the-art natural language processing models. Moreover, in comparison to the original scheme of contextual perturbation CLARE, SASSP has yielded a higher attack success rate and lower word perturbation rate

## 1 Introduction

Adversarial attacks in natural language processing (NLP) have become a pivotal area of research, exposing the vulnerabilities of machine learning models to small, barely noticeable changes in input texts. These changes are crafted to mislead models into incorrect predictions while still appearing normal and unchanged to human readers. Adversarial text attacks hold significant importance for the robustness and security of NLP systems used in critical areas such as sentiment analysis, machine translation, and automated customer support. The development of adversarial attacks traces back to influential studies[1-2], revealing how deep neural networks (DNNs) can be influenced by imperceptible changes. These findings spurred extensive research, resulting in numerous attack methods and corresponding defense strategies.

Initial adversarial text attack methods often involved basic alterations like character-level changes (e.g., typos, misspellings) or synonym substitutions. Although these methods were sometimes effective, they frequently disrupted the text's naturalness and readability. Furthermore, they struggled to maintain the original text's semantic meaning, which is crucial for ensuring the adversarial example remains coherent and contextually appropriate.

More advanced techniques have since emerged, focusing on word- and sentence-level perturbations. These methods aim to make more meaningful changes that preserve the overall readability and semantic content of the text. Techniques such as word embeddings and masked language models (MLMs) have been used to generate candidate perturbations that are contextually plausible. These advancements have led to the development of more effective adversarial examples.

A significant advancement in this field is the use of contextual information to guide perturbations. By leveraging the context provided by surrounding words, these methods can generate perturbations that are more likely to maintain the text's coherence and meaning. Among the various methods leveraging contextual information, the CLARE (Contextualized Adversarial Examples) method proposed by Li et al. is particularly noteworthy [3]. CLARE generates adversarial examples by perturbing words in a way that considers the broader context of the text. This is achieved by using masked language models to predict potential substitutes for target words, ensuring that these substitutes are contextually appropriate.

CLARE uses MLMs to ensure that word substitutions are contextually plausible, maintaining local coherence. Leveraging context helps preserve the overall meaning of the text better than earlier methods. It effectively misleads various state-of-the-art NLP models, demonstrating its utility as an adversarial attack method. However, CLARE primarily relies on gradient-based methods to identify salient words for perturbation. This can overlook contextually important words that are not highlighted by gradients alone.

*Objective:* Our goal is to address the weaknesses of the CLARE method by enhancing its approach to word selection and semantic consistency verification. We introduce "Saliency Attention and Semantic Similarity Driven Adversarial Perturbation" (SASSP), a novel method that integrates saliency, attention, and semantic similarity to improve the generation of adversarial examples.

*Contributions:* Our contributions are the following. First, we develop a combined saliency-attention score to identify candidate words for perturbation. This score combines gradient-based saliency with attention scores from transformer models, ensuring that perturbations focus on words that are both impactful and contextually significant. Second, we implement a comprehensive semantic consistency verification process that includes embedding-based similarity, and paraphrase detection. This makes sure that the perturbed text remains semantically identical to the source and preserves the roles and relationships of entities within the text.

By addressing the limitations of the CLARE scheme, our SASSP method aims to advance the field of textual adversarial attacks. The integration of saliency, attention, and semantic similarity ensures that our generated adversarial examples maintain high semantic fidelity while effectively deceiving state-of-the-art NLP models.

The remainder of this paper is as follows. In Section 2, we present some related work in the field of adversarial attacks on text-based models, focusing on existing methods and their limitations. Section 3 presents the steps involved in the operations of the contextual perturbation in CLARE. In Section 4, we present our proposed SASSP scheme, detailing the integration of a robust word selection strategy for perturbation and a semantic similarity-checking module with the CLARE framework. In Section 5, we present experimental results demonstrating the effectiveness and robustness of SASSP in comparison to the CLARE scheme. Section 6 concludes the paper and outlines avenues for further research.

## 2 Related Work

Carlini & Wagner present a method to evaluate the robustness of neural networks, particularly in the context of security and privacy [4]. They propose a comprehensive framework to generate adversarial examples, which are inputs specifically crafted to fool the neural network into making incorrect predictions.

Jia & Liang propose a method to generate adversarial examples specifically designed for assessing reading comprehension systems [5]. Their approach involves making subtle alterations to input passages and questions, aiming to provoke misinterpretations or incorrect answers from the comprehension models. Through systematic modifications, the authors illustrate the susceptibility of existing reading comprehension systems to adversarial attacks.

Liu & Lane propose a novel approach to improve the robustness and performance of task-oriented dialog systems using adversarial learning techniques [6]. The authors introduce an adversarial training framework where a generator, which creates dialog responses, and a discriminator, which evaluates the quality and relevance of these responses, are trained simultaneously. This approach helps the dialog model to generate more natural and contextually appropriate responses.

Wei et al. present TextBugger, a technique aimed at generating adversarial text for real-world applications [7]. TextBugger focuses on making subtle alterations to input text, which result in misclassifications or erroneous behavior in various natural language processing (NLP) systems.

Ren et al. introduce a method to generate natural language adversarial examples by leveraging probability-weighted word saliency [8]. Their approach focuses on identifying influential words in input text using a probabilistic weighting scheme to generate effective adversarial examples. By targeting these salient words, the authors demonstrate the ability to induce misclassifications or alter the semantic meaning of the text, thus challenging the robustness of NLP models.

Yang et al. propose a novel method for generating adversarial text attacks [9]. This approach focuses on altering text inputs to deceive NLP models while keeping the semantic meaning intact. By ensuring that the altered text maintains its original meaning, the proposed method aims to create adversarial examples that are indistinguishable from genuine inputs to human readers.

Liu et al. present a method to generate transferable adversarial examples with minimal changes [10]. Their approach aims to create adversarial examples that can fool multiple machine learning models while making the smallest possible alterations to the input. By minimizing changes, the generated adversarial examples are more likely to be transferable across different models and datasets.

Waghela et al. propose an enhanced method for generating adversarial examples targeted at text classification models [11]. Their approach builds upon word saliency-based techniques to identify and manipulate crucial words in text inputs, effectively altering their semantic meaning. By modifying the word saliency computation, the authors aim to improve the efficiency and effectiveness of the adversarial attack.

In addition to the schemes discussed above, there are several works in the literature involving word substitution [12], word insertion [13], word swapping [14], phrase perturbation [15], sentence perturbation [16], and contextual perturbation [3].

Despite these advancements, existing methods still face challenges in generating adversarial examples that are both effective and imperceptible. Many methods focus solely on maximizing the model's prediction error without considering the semantic similarity or grammatical correctness of the adversarial examples. As a result, the generated examples may be nonsensical or linguistically unnatural, limiting their practical utility in real-world applications.

## 3   The CLARE Model

Adversarial attacks in NLP aim to mislead models by making subtle modifications to input texts. These modifications must be imperceptible to humans but significant enough to cause the model to make incorrect predictions. Li et al introduced the contextualized adversarial example (CLARE) model, a novel approach that uses pre-trained language models to generate context-aware and semantically coherent adversarial examples [3]. Contextual perturbation focuses on creating adversarial examples that maintain the original text's fluency and meaning while effectively misleading text classification models. This is achieved through several steps involving mathematical and computational techniques discussed briefly in the following.

*(1) Selection of target words:* The words in the text are identified that, when altered, can significantly impact the model's predictive power. These target words are chosen based on their importance to the model's output. The importance is determined using gradient-based saliency scores. Given an input text $X = \{x_1, x_2, \ldots, x_n\}$, where $x_i$s are the words, and a model $f(X)$ that predicts a label $y$, the gradient of the loss $L(y, f(X))$ is computed concerning each word embedding $e_i$ using (1).

$$\frac{\partial L(y, f(X))}{\partial e_i} \tag{1}$$

The saliency score $s_i$ for each word $x_i$ is given by the norm of the gradient in (2).

$$s_i = \left\|\frac{\partial L(y, f(X))}{\partial e_i}\right\| \tag{2}$$

The top-$k$ words are selected with the highest saliency scores as target words for perturbation.

*(2) Contexualized generation:* The model then generates perturbation by considering the entire context of the text using a masked language model (MLM) such as BERT. For each target word $x_t$, a masked version of the text is created as in (3).

$$X_{[MASK]} = \{x_1, \ldots, x_{t-1}, [MASK], x_{t+1}, \ldots, x_n\} \tag{3}$$

An MLM is used to predict a set of candidate words $\{c_1, c_2, \ldots, c_k\}$ for the masked position. The probability of each candidate $c_i$ is given by (4).

$$P(c_i | X_{[MASK]}) = MLM(X_{[MASK]}) \tag{4}$$

*(3) Ranking and filtering:* The generated candidates are ranked based on their likelihood and their potential to fool the model. For this purpose, each candidate word $c_i$ is evaluated by its likelihood and its impact on the model's prediction. The combined score for each candidate is derived using (5).

$$score(c_i) = \alpha P(c_i | X_{[MASK]}) - \beta |f(X_{c_i}) - f(X)| \tag{5}$$

In (5), α and β are the weighting factors, and $X_{c_i}$ is the text with $x_t$ replaced by $c_i$. Finally, the candidates $\{c_1, c_2, \ldots, and\ c_k\}$ are ranked based on their scores, and the top candidate is selected.

*(4) Adversarial example generation:* The selected target words are replaced with the top-ranked perturbations. The selected target word $x_t$ in $X$ with top-ranked candidate $c_{top}$ to create the adversarial text $X'$ in (6).

$$X' = \{x_1, \ldots, x_{t-1}, c_{top}, x_{t+1}, \ldots, x_n\} \tag{6}$$

If necessary, the process is repeated iteratively for multiple target words until the adversarial example $X'$ successfully misleads the model $f$.

However, the overall effectiveness of the CLARE method depends on several key factors such as (i) quality and capacity of the pre-trained language model used to generate the perturbation, (ii) contextual embedding ability of the language model, (iii) robustness of the word importance score computation, (iv) synonym replacement quality, (v) semantic similarity checks, (vi) grammatical correctness of the adversarial texts, (vii) task relevance, and (viii) regularization techniques used.

While contextual perturbation used in CLARE is a sophisticated and very effective adversarial text attack, there are a few potential weaknesses in the scheme. Four notable such shortcomings are (i) sensitivity to the choice of target words, (ii) reliance on masked language models, (iii) high computational complexity, and (iv) limitations in maintaining semantic resemblance between the adversarial and the source texts. We focus on points (i) and (iv) and propose an improved and more robust attack strategy.

In Section 4, our proposition integrating a more robust word selection approach and a semantic similarity check with the CLARE scheme is presented. The effectiveness of the proposed scheme is shown in Section 5.

## 4 The Proposed Scheme of Adversarial Perturbation

While the process of generating adversarial perturbation in text using contextual perturbation is a very effective adversarial text attack, as mentioned in Section 3, it involves two key challenges: (i) selecting the right words to perturb and (ii) ensuring that the adversarial text maintains semantic resemblance to the source text. We first explain these two challenges and explore the problems associated with them.

*Word selection in contextual perturbation:* Selecting the appropriate words to perturb is crucial because perturbing the wrong words can lead to either ineffective adversarial examples or examples that are easily detectable. The challenge lies in identifying words that are significant to the model's prediction while ensuring that the perturbation does not make the text semantically nonsensical or unnatural.

Saliency scores indicate how much each word contributes to the model's prediction by measuring the gradient of the loss for the word embeddings. However, these scores can sometimes highlight irrelevant words if the model's gradients are noisy or if the loss landscape is complex. On the other hand, attention scores from transformer models show which words the model focuses on, but these scores alone might not fully capture the importance of each word to the final prediction, especially in tasks where attention is distributed widely across many words. Combining saliency and attention scores requires setting appropriate weighting factors. Incorrect weights can either overemphasize or underemphasize certain words, leading to suboptimal perturbations. Moreover, setting a dynamic threshold for selecting words to perturb is tricky. A threshold that is too low might select too many words, diluting the attack's focus, while a threshold that is too high might miss important words. Finally, perturbing words without considering their contextual roles can lead to grammatically incorrect or contextually irrelevant modifications. For example, changing a noun to an unrelated noun can confuse the reader and reduce the attack's stealthiness.

*Maintaining semantic similarity:* Maintaining semantic similarity ensures that the perturbed text conveys the same meaning as the original text. This is crucial for the perturbation to remain undetected by human readers and automated defenses while still fooling the target model.

Perturbations can inadvertently change the meaning of the text. For example, altering a critical word can alter the sentiment or introduce ambiguity, making the perturbed text semantically different from the original. Even if a word is contextually appropriate, its substitution might not fit semantically with the surrounding words, leading to a text that feels off to human readers. Maintaining role consistency is a more complex task. Semantic Role Labeling (SRL) helps ensure that key entities and actions remain consistent in the perturbed text. However, inaccuracies in SRL models can lead to missed or incorrect roles, making it difficult to maintain consistency. Moreover, comparing semantic roles between the original and perturbed texts is non-trivial, as subtle differences in role assignments can lead to significant meaning change.

Making sure that the replaced words closely match the original ones aids in preserving the theme and tone of the text. However, lexical similarity alone doesn't guarantee semantic similarity. Again, the perturbation must preserve grammatical correctness. Perturbing words without regard to syntax can lead to sentences that are grammatically incorrect and thus easily detectable as adversarial examples.

Our proposed *Saliency Attention and Semantic Similarity Driven adversarial Perturbation* (SASSP) involves the following two additional components.

*(1) Robust word selection strategy:* To address the problem in word selection, in the proposed SASSP scheme, a combined approach leveraging both saliency and attention scores is used. While the saliency of a word $x_i, s_i$ is computed using (2), the attention score $a_i$ for the word $x_i$ is derived using (7).

$$a_i = \frac{1}{L*H} \sum_{l=1}^{L} \sum_{h=1}^{H} A_i^{(l,h)} \qquad (7)$$

In (7), $a_i$ is the attention score for word $x_i$, $A_i^{(l,h)}$ is the attention weight for word $x_i$ in layer $l$ and head $h$, $L$ is the number of layers, and $H$ is the number of heads.

The saliency score *is* $s_i$ computed using (1), and the attention score $a_i$ computed using (7) for each word $x_i$ is then combined into a composite score $c_i$ using (8).

$$c_i = \alpha s_i + \beta a_i \qquad (8)$$

In (8), $c_i$ is the combined score for word $x_i$, and α and β are the weighting factors.

In the proposed SASSP scheme, an adaptive thresholding approach has been used. A threshold is computed using (9).

$$threshold = \gamma * \max(c_i) \qquad (9)$$

In (9), $\gamma$ is a tunable parameter, and target words are then selected based on (10).

$$T = \{x_i = c_i \geq threshold\} \qquad (10)$$

In other words, those words whose combined saliency and attention scores exceed the threshold value $T$, are the candidate words for perturbation.

The effectiveness of using gradient-based saliency scores depends on how well the gradients reflect the importance of each word. The norm of the gradient vector for each word's embedding represents its influence on the model's loss. Moreover, the gradients should be computed concerning the most sensitive part of the model's prediction. This requires careful consideration of the loss function.

The attention scores from transformer models provide insights into which words the model focuses on during prediction. The effectiveness of using attention scores depends on two factors: (i) aggregation strategy and (ii) relevance to the target model. The method of aggregating attention scores across layers and heads should retain important contextual information. This involves calculating the mean or weighted sum of attention scores. Moreover, the attention scores should be relevant to the specific task and domain of the target model. This may require fine-tuning the model to align its attention mechanism with the target task.

Combining saliency and attention scores effectively requires balancing their contribution by choosing appropriate values for the weights α and β. These factors should be ideally optimized through cross-validation.

An adaptive threshold for selecting target words ensures that only the most impactful words are perturbed. Using a dynamic threshold based on the distribution of combined scores ensures that the selection is adaptive to different texts and contexts. The threshold value directly influences the number of words selected for perturbation and their impact on the model's prediction.

*(2) Semantic Consistency Preservation:* To ensure semantic consistency, our proposed scheme SASSP incorporates multiple layers of verification. These layers include (i) *embedding-based semantic similarity* and (ii) *paraphrase detection*.

The objective of the embedding-based semantic similarity detection module is to ensure that the perturbed text remains semantically similar to the original text. Using the sentence embedding model of BERT the original and perturbed texts are encoded into high-dimensional vectors. The cosine similarity between these vectors is then computed using (11).

$$sim(X, X') = \frac{E(X).E(X')}{\|E(X)\|\|E(X')\|} \quad (11)$$

In (11), $E(X)$ and $E(X')$ are embeddings of the original and perturbed texts. Only those words yielding semantic similarities above a pre-set threshold are selected.

The paraphrase detection model evaluates the probability that the perturbed text is a paraphrase of the original text. The pre-trained Sentence Transformers Models from the Sentence Transformers library are used for paraphrase detection and scoring the similarity. Those words which pass through the cosine similarity test are passed through the paraphrase detection module. Only those words that exceed a pre-set threshold of the paraphrase score are finally selected for perturbation.

Finally, we outline the sequential steps involved in the proposed scheme.

*(a) Input text and model preparation:* The original text $X$ is taken as the input and the target model $f$ is identified that is targeted to be attacked.

*(b) Saliency score computation:* The gradient of the loss $L$ is computed concerning each word embedding $e_i$ in the text. The saliency score $s_i$ for each word is computed as the norm of the gradient.

*(c) Attention score computation:* The attention weights $A_i^{(l,h)}$ are extracted from the target model across all layers $L$ and heads $H$. The attention score for each word is computed as the average attention weight.

*(d) Combining saliency and attention scores:* A combined score $c_i$ for each word $x_i$ is computed by weighting (α and β), and summing the saliency and attention scores.

*(e) Selecting words for perturbation:* An adaptive threshold is used for selecting target words based on their combined scores. Words with their composite scores greater than the threshold are used for perturbation.

*(f) Generating perturbations:* For each selected word, a mask language model is used to generate contextually appropriate substitutions. A list of candidate words for each target word is generated that fits the context of the original sentence.

*(g) Evaluating semantic similarity:* The semantic similarity between the original text and each candidate perturbed text is computed using sentence embeddings. Then, a paraphrase detection model is used to evaluate the probability that the adversarial text paraphrases the original text. Perturbations that meet a certain threshold of semantic similarity and paraphrase probability are finally accepted as adversarial texts.

## 5   Performance Results

The performance of our proposed SASSP scheme is compared with that of CLARE. As in the evaluation of CLARE, the distilled version of RoBERTAa [17] is used as the masked language model. Distill-RoBERTa is a distilled variant of the RoBERTa-base model, adhering to the same training methodology as DistillBERT.

The semantic similarity between the original text and adversarial text is computed based on the *universal sentence encoder* (USE) [18]. USE converts sentences into fixed-length vectors, which capture the semantic meaning of the sentences. To compute the similarity of their corresponding embedding vectors is calculated. Cosine similarity measures the cosine of the angle between two vectors, which indicates how similar the vectors (and thus the sentences) are.

The open-source implementation of CLARE is used for evaluating CLARE. To evaluate how our proposed SASSP attack performs in comparison to CLARE, we utilize the following four datasets that were used in evaluating CLARE.

*Yelp:* This dataset is used for sentiment analysis and NLP tasks. It consists of reviews covering businesses like restaurants, cafes, and hotels. Each review includes a user-given star rating, typically from 1 to 5 stars. Ratings of 4 and 5 stars are labeled as positive, while 1 and 2 stars are considered negative. The dataset contains 560,000 training samples and 38,000 test samples, with polarity class 2 representing positive sentiment and class 1 representing negative sentiment. The dataset is employed for polarity classification tasks. The average length of the reviews is 130 words and the accuracy of the target model, DistillRoBERTa for this dataset is 95.9% in the absence of any adversarial attack.

*AG News:* This dataset comprises news articles from the AG's corpus of online news articles. The articles are categorized into four groups: world, sports, business, and science & technology. It includes 120,000 training samples and 7,600 test samples. The average length of the reviews is 46 words and the accuracy of the victim model for this dataset is 95.0% in the absence of attacks.

*MNLI:* This dataset is used for textual entailment tasks. It includes a total of 393,702 pairs for sentence pairs, with 392, 702 pairs for training, 20,000 for validation, and 20,000 for testing. MNLI features sentence pairs from various genres such as fiction, government, editorial articles, telephone conversations, and travel. The accuracy of the victim model on this dataset is 84.3% in the absence of any attacks.

*QNLI:* This dataset originates from the Stanford Question Answering Dataset. Its purpose is to assess whether a context sentence includes the answer to a given question. Each example in the dataset consists of a question, a context sentence from the corresponding passage, and a label indicating whether the sentence includes the answer to the question. The labels are "entailment" (if the context sentence includes the answer) and "not entailment" (if the context sentence does not contain the answer). The training set contains 104,743 entries, with the validation and test sets each comprising 5,463 entries. The accuracy of the victim for this dataset in the absence of adversarial attacks is 91.4%.

*Metrics for evaluation:* The performance of SASSP is compared with CLARE on the above 4 datasets on the following metrics.

*Attack Success Rate (ASR):* It refers to the proportion of adversarial samples that effectively compromise the integrity of the victim model. An attack scheme with a higher ASR is more effective.

*Perplexity (PER):* The perplexity of an attack model refers to the level of uncertainty or unpredictability exhibited by the model when generating adversarial samples. A lower perplexity indicates that the model is more confident and consistent in generating effective attacks. The computation of perplexity is done following the approach discussed in [19]. A higher PER implies a more effective model.

*Word Manipulation Rate (WMR):* It quantifies the percentage of altered tokens within a text. Each alteration, whether through replacement or insertion, contributes to the modification count. A single token modification is attributed to each merge action, except when two merged tokens are retrained, in which case it is considered as modifying two tokens. An attack scheme with less WMR and high ASR is a more effective scheme.

*Syntactic Error (SYE):* It quantifies the total count of newly introduced grammatical errors in a successfully generated adversarial example when compared to the original text. The scheme presented by Naber et al. is used for computing the value of this metric [20]. An attack scheme that yields a lower value of SYE for its adversarial text is a more effective one.

*Semantic Similarity (SES):* Semantic similarity refers to the degree of likeness or resemblance in meaning between two pieces of text. It captures the extent to which two linguistic expressions covey similar concepts, ideas, or semantics, regardless of their surface form or syntactic structure. The semantic similarity is computed using universal sentence encoding [18]. The higher the value for SES, the more efficient the attack scheme.

Table 1. The comparative performance of CLARE and SASSP schemes on Yelp dataset

| Model | ASR | PER | WMR | SYE | SES |
|---|---|---|---|---|---|
| CLARE | 79.7 | 83.5 | 10.3 | 0.25 | 0.78 |
| SASSP | **82.6** | **74.2** | **8.7** | **0.25** | **0.85** |

Table 2. The comparative performance of CLARE and SASSP schemes on AG News dataset

| Model | ASR | PER | WMR | SYE | SES |
|---|---|---|---|---|---|
| CLARE | 79.1 | 86.0 | 6.1 | 0.17 | 0.76 |
| SASSP | **83.4** | **76.2** | **3.4** | **0.15** | **0.85** |

Table 3. The comparative performance of CLARE and SASSP schemes on MNLI dataset

| Model | ASR | PER | WMR | SYE | SES |
|---|---|---|---|---|---|
| CLARE | 88.1 | 82.7 | 7.5 | **0.02** | 0.82 |
| SASSP | **93.6** | **73.2** | **3.3** | **0.02** | **0.87** |

**Table 4.** The comparative performance of CLARE and SASSP schemes on QNLI dataset

| Model | ASR | PER | WMR | SYE | SES |
|---|---|---|---|---|---|
| CLARE | 83.8 | 76.7 | 11.8 | 0.01 | 0.78 |
| SASSP | **93.2** | **71.2** | **5.8** | **0.01** | **0.86** |

Tables 1- 4 present the comparative performance results of our proposed SASSP scheme and CLARE. The results are based on 1000 test records from the respective datasets. For an attack scheme to be more effective and efficient it must have higher values for ASR and SES and lower values for PER, WMR, and SYE. It is observed from Tables 1-4 that SASSP outperforms CLARE on all four datasets and for all five metrics. It may be noted that the PER, WMR, SYE, and SES values presented in Tables 1-4 are the average values computed for the adversarial samples that were successful in attacking the victim model.

**Table 5.** CLARE and SASSP compared with different MLMs on the AG News dataset

| MLM | Attack | ASR | PER | WMR | SYE | SES |
|---|---|---|---|---|---|---|
| Distill-RoBERTa | CLARE | 79.1 | 86.0 | 6.1 | 0.17 | 0.76 |
|  | SASSP | **83.4** | **76.2** | **3.4** | **0.15** | **0.85** |
| Base-RoBERTa | CLARE | 79.3 | 88.9 | 6.3 | 0.21 | 0.75 |
|  | SASSP | **81.7** | **77.2** | **4.7** | **0.17** | **0.85** |
| Base-BERT | CLARE | 78.4 | 95.2 | 8.3 | 0.23 | 0.71 |
|  | SASSP | **81.4** | **78.5** | **6.5** | **0.19** | **0.85** |

Finally, the performances of SASSP and CLARE are compared on three different MLMs, Distill-RoBERTa, Base-RoBERTa, and BASE-BERT on the AG News dataset. The results are presented in Table 5. It is observed that SASSP outperforms CLARE on all five metrics considered in the evaluation framework. It is also observed that among the three MLMs, while Distill-RoBERTa yielded the best results for both attack schemes, the Base-BERT exhibited the worst results for them.

## 5 Conclusion

This paper introduced a new adversarial attack technique, Saliency Attention, and Semantic Similarity driven adversarial Perturbation (SASSP). The proposed scheme exhibits better performance than the contextual perturbation-based adversarial method CLARE across multiple evaluation metrics. SASSP achieves higher attack success rates, lower word manipulation rates, and lower perplexity scores, indicating its effectiveness in generating adversarial examples with minimal modifications while maximizing deception. Additionally, SASSP produces adversarial texts with higher syntactic correctness and semantic similarity to the original text, preserving both readability and meaning. Moving forward, further research into the interpretability and robustness of adversarial examples generated by SASSP could provide valuable insights into model vulnerabilities and defense mechanisms.

# References

1. Szegedy, C., Zaremba, W., Sutskever, I., Bruma, J., Erhan, D., Goodfellow, I.J., and Fergus, R: Intriguing Properties of Neural Networks. In: Proc. of ICLR, Poster Track (2014)
2. Goodfellow, I.J., Shlens, J., and Szegedy, C.: Explaining and Harnessing Adversarial Examples. In: Proc. of ICLR, Poster Track (2015)
3. Li, D., Zhang, Y., Peng, H., Chen, L., Brockett, C., Sun, M-T., Dolan, B: Contextualized Perturbation for Textual Adversarial Attack. In Proc. of NAACL, pp 5053-5069 (2021)
4. Carlini, N., Wagner, D.: Towards Evaluating the Robustness of Neural Networks. In: Proc of IEEE Symposium on Security and Privacy, pp 39-57 (2017)
5. Jia, R., Liang, P.: Adversarial Examples for Evaluating Reading Comprehension Systems. In: Proc. of Conf. on Emp. Methods in Nat. Lang. Processing, pp 2021-2031 (2017)
6. Liu, B., Lane, I.: Adversarial Learning of Task-Oriented Neural Dialog Models. In: Proc. of the 19[th] Annual SIGdial Meeting on Discourse and Dialogue, pp 350-359 (2018)
7. Wei, J., Zou, K., Cao, T., Chen, Z., Huang, Y.: TextBugger: Generating Adversarial Text against Real-World Applications. In: Proc. of ACM SIGSAC, pp 1969-1986 (2019)
8. Ren, S., et al.: Generating Natural Language Adversarial Examples through Probability Weighted Word Saliency. In: Proc. of the 57[th] Conf of ACL pp 1085-1097 (2019)
9. Yang, X., Gong, Y., Liu, W., Baily, J., Tao, D., Liu, W.: Semantic-Preserving Adversarial Text Attacks. IEEE Trans on Sustainable Computing, Vol 8, No 4, pp 583-595 (2023)
10. Liu, F., Zhang, C., Zhang, H.: Towards Transferable Unrestricted Adversarial Examples with Minimum Changes. In: Proc. of IEEE SaTML, pp 327-338 (2023)
11. Waghela, H., Rakshit, S. and Sen, J.: A Modified Word Saliency-Based Adversarial Attack on Text Classification Models. In Proc. of ICCIDA, *arXiv:2403.11297* (2024)
12. Waghela, H., Sen, J., and Rakshit, S.: Enhancing Adversarial Text Attacks on BERT Models with Projected Gradient Descent. In: Proc. of IEEE ASIANCON Pune, India. (2024)
13. Ni, M., Sun, Z., Liu, W.: Frauds Bargain Attack: Generating Adversarial Text Samples via Word Manipulation Process. IEEE Trans on Knowledge and Data Engineering (2024)
14. Liu, H. et al.: Textual Adversarial Attacks by Exchanging Text-Self Words. Int. J. of Intelligent System, Vol 37, No 12, pp 12212-12234 (2022)
15. Lei, Y., Cao, Y., Li, D., Zhou, T., Fang, M., and Pechenizkiy, M.: Phrase-Level Textual Adversarial Attack with Label Preservation. Findings of the ACL, pp 1095-1112 (2022)
16. Li, A., Zhang, F., Li, S., Chen, T., Su, P., Wang, H.: Efficiently Generating Sentence-Level Adversarial Examples with Seq2Seq Stacked Autoencoder. Exert System with Applications, Vol 213, Part C, Art ID: 119170 (2021)
17. Sanh, V., Debut, L., Chaumond, J., Wolf, T.: DistilBERT, A Distilled Version of BERT: Smaller, Faster, Cheaper and Lighter. In: Proc. of 5[th] Workshop of NeurIPS (019)
18. Cer, D., et al.: Universal Sentence Encoder. arXiv:1803.11175 (2018)
19. Radford, A., Wu, J., Child, R., Luan, D., Amodei, D., Sutskever, I.: Language Model are Unsupervised Multitask Learners. OpenAI Blog Vol 1, No 8 (2019)
20. Naber, D. et al.: A Rule-Based Style and Grammar Checker. Citeseer, (2003)